\documentclass{article}

\usepackage{arxiv}

\usepackage[utf8]{inputenc} 
\usepackage[T1]{fontenc}    
\usepackage{hyperref}       
\usepackage{url}            
\usepackage{booktabs}       
\usepackage{amsfonts}       
\usepackage{nicefrac}       
\usepackage{microtype}      
\usepackage{lipsum}
\usepackage{graphicx}
\usepackage{multirow}
\usepackage{subcaption}
\graphicspath{ {./images/} }

\title{Transformer-Based Deep Learning Model for Stock Price Prediction: A Case Study on Bangladesh Stock Market}

\author{
 Tashreef Muhammad \\
  Department of Computer Science and Engineering \\
  Ahsanullah University of Science and Technology \\
  Dhaka, Bangladesh \\
  \texttt{tashreef.cse@aust.edu} \\
   \And
 Anika Bintee Aftab \\
  Department of Computer Science and Engineering \\
  Ahsanullah University of Science and Technology \\
  Dhaka, Bangladesh \\
  \texttt{anikaaftab.cse@aust.edu} \\
  \And
 Md. Mainul Ahsan \\
  Department of Computer Science and Engineering \\
  Ahsanullah University of Science and Technology \\
  Dhaka, Bangladesh \\
  \texttt{mainulahsanswaad123@gmail.com} \\
   \And
   Maishameem Meherin Muhu \\
  Department of Computer Science and Engineering \\
  Ahsanullah University of Science and Technology \\
  Dhaka, Bangladesh \\
   \texttt{maishamuhu111@gmail.com} \\
   \And
   Muhammad Ibrahim \\
   Department of Computer Science and Engineering \\
   University of Dhaka \\
   Dhaka, Bangladesh \\
   \texttt{ibrahim313@du.ac.bd} \\
   \And
   Shahidul Islam Khan \\
   Department of Computer Science and Engineering \\
   International Islamic University Chittagong \\
   Chittagong, Bangladesh \\
   \texttt{shahid@iiuc.ac.bd} \\
   \And
   Mohammad Shafiul Alam \\
   Department of Computer Science and Engineering \\
   Ahsanullah University of Science and Technology \\
   Dhaka, Bangladesh \\
   \texttt{shafiul.cse@aust.edu} \\
}

\begin{document}
\maketitle
\begin{abstract}
In modern capital market the price of a stock is often considered to be highly volatile and unpredictable because of various social, financial, political and other dynamic factors. With calculated and thoughtful investment, stock market can ensure a handsome profit with minimal capital investment, while incorrect prediction can easily bring catastrophic financial loss to the investors. This paper introduces the application of a recently introduced machine learning model --- the Transformer model, to predict the future price of stocks of Dhaka Stock Exchange (DSE), the leading stock exchange in Bangladesh. The transformer model has been widely leveraged for natural language processing and computer vision tasks, but, to the best of our knowledge, has never been used for stock price prediction task at DSE. Recently the introduction of \emph{time2vec} encoding to represent the time series features has made it possible to employ the transformer model for the stock price prediction. This paper concentrates on the application of transformer-based model to predict the price movement of eight specific stocks listed in DSE based on their historical daily and weekly data. Our experiments demonstrate promising results and acceptable root mean squared error on most of the stocks.
\end{abstract}


\section{Introduction}\label{sec: introduction}

Stock market is usually considered to be unpredictable for its dynamic behavior due to its dependence on many external and/or environmental factors. However, the investors who are expert in domain knowledge believe that in many situations, the nature of the stock market is not completely unpredictable, rather a thoughtful investment may foresee future events to some extent. Besides, the major market-altering events are not very frequent either. Hence, instead of manually predicting future stock market events such as price of a stock, computational scientists and practitioners have exerted effort to automatically predict stock market movement and to build a robust, profitable mechanism for years \cite{SurveyBio, obthong_survey_2020}.

\subsection{Attention-Based Recurrent Neural Networks}

Among a plethora of computational techniques, machine learning models yield highly effective predictions of future events provided they are fed with sufficient and relevant historical data. Artificial neural network is one of the most powerful machine learning techniques that has been found to be tremendously effective in many domains. Transformer is a highly effective attention based neural network model that is being heavily used for Natural Language Processing (NLP) tasks since its introduction in 2017 \cite{vaswani2017attention-transformer}. The model is particularly popular  for its high accuracy on sequence data modelling. The strength of transformer model are mainly due to for its unique use of the concepts like attention and encoder-decoder architecture. 

Another recent development, namely the time2vec encoding mechanism \cite{mehran2019time2vec} has enabled the transformer model to be used in  time series analysis problems such as stock price prediction. A time series is a sequence of numerical data points in successive order. However, very few existing works have been found (e.g.: \cite{transformer-timeembedding}) in our literature survey where transformer models integrated with time2vec have been employed for stock price prediction (as detailed in Section~\ref{sec: related work}). Moreover, to the best of our knowledge, such an approach has never been used on the stock price data of the Dhaka Stock Exchange (DSE), the leading stock market of Bangladesh.

\subsection{Motivation and Contribution}

As mentioned above, attention based recurrent neural network models are known to be very effective in sequence-to-sequence modelling. Of these, transformer based models top this list. These models can be adapted to time series data such as stock market data. Being the leading stock market of Bangladesh (which hosts a staggering figure of 180 million population), Dhaka Stock Exchange has a tremendous amount of digital data but has not been so far subject to rigorous machine learning investigation. We, therefore, think that there is a high potential for machine learning models in general, and attention based recurrent neural network models in particular, for Bangladesh stock price prediction. This study is an endeavour to explore this potential.

The main contributions of our paper are: (1) developing a transformer based machine learning model for DSE stock price prediction, (2) incorporating time2vec embedding in our model, and (3) yielding satisfactory performance on DSE data, thereby unravelling potential benefit of using machine learning models for Bangladesh stock market forecasting. 

The rest of the paper is organized as follows. In Section~\ref{sec: related work} we discuss the relevant research works. Section~\ref{sec: methodology} describes our methodology that details data collection process and the learning model. Section~\ref{sec: result} discusses the experimental findings. Section~\ref{sec: conclusion} concludes the paper.

\section{Related Work}
\label{sec: related work}

Research related to stock price prediction using prediction techniques like neural networks has been ongoing for more than thirty years \cite{schoneburg1990stock}. However, there is still room for investigation as the problem of predicting stock price is a non-trivial one. Below we discuss the relevant existing works. 

Wamkaya et al. \cite{wanjawa2014ann} build a prototype using Artificial Neural Network (ANN) and tested on three New York Stock Exchange stocks. Related to this idea, several research works have been conducted to predict stock price using Convolutional Neural Networks (CNNs) as well \cite{tsantekidis2017forecasting, gudelek2017deep, selvin2017stock, hiransha2018nse, kim2019forecasting, chen2018stock} with reasonably good performance. Since stock prices possess properties of sequence data, both vanilla Recurrent Neural Networks (RNN) and Long-Short Term Memory (LSTM) models have been utilized \cite{selvin2017stock, hiransha2018nse, kim2019forecasting}.  Transformer based models for stock price prediction is also picking up pace. It has already been used to forecasting S\&P volatility \cite{ramos2021multi}. Transformer models have also been used on natural language data collected from social media related to stock price forecasting \cite{liu2019transformer}. 

On another front, a number of researchers have used a variety of Artificial Intelligence (AI) techniques in stock price prediction \cite{obthong_survey_2020}. Especially the so-called evolutionary and bio-inspired algorithms lead the deployment of meta-heuristics and AI-based techniques such as Genetic Algorithm, Artificial Bee Colony, Ant Colony, Fish Swarm optimization, Particle Swarm Optimization and the like \cite{SurveyBio}. Techniques of time series analysis like Box Jenkins method have also been used in some studies \cite{PSO_Box}.

Now we concentrate on the existing few works on DSE data. Using the name ``Dhaka Stock Exchange'', DSE officially started its journey on $16^{th}$ September, 1986 \cite{dse-found}. During the literature survey, only a few research works have been found that attempts to predict DSE stocks' movements based on their historical data. Kamruzzaman et al. \cite{kamruzzaman2017modeling} published a study that uses Box-Jenkins methodology and applied Autoregressive Integrated Moving Average (ARIMA) to find interval forecasts of market return of DSE with 95\% confidence level. Maksuda et al. \cite{rubi2019forecasting} predicted the DSE Broad Index (DSEX) using a multi-layer feed-forward neural network and report satisfactory performance.

Mujibur et al. \cite{BD_timeseriesforecast} deployed ARIMA, an artificial neural network, linear model, Holt-Winters model, and Holt-Winters exponential smoothing model on as many as 35 stocks of DSE and report the artificial neural network to perform relatively better compared to the other techniques.

However, from our literature survey, as discussed above, we have found no research study that employs the Transformer moddel along with the time2vec \cite{mehran2019time2vec} encoding on DSE data. This paper is aimed at bridging this research gap.

\section{Proposed Methodology}
\label{sec: methodology}

This section discusses the collection and preparation of DSE data and the proposed application of the transformer model for future price prediction.

\subsection{Data Collection and Preparation}

The initial data have been collected from \textbf{\textit{Amarstock}}\footnote{\url{https://www.amarstock.com/}}, a DSE data providing website, of the time frame October, 2012 till December, 2020. These data consist of seven features, which are -- the Trading Code, Date, Opening Price, High Price, Low Price, Closing Price and Volume. Based on the Trading Code, the target company's data are extracted from the entire dataset to train the transformer model, which means a separate model is trained for each different stock data. The models are based on End Of Day (EOD) data, which means all the features are collected once per day per stock. From the daily EOD data, the weekly data are prepared which define the state of the stock per week. Conventionally, the DSE week consists of five working days (Monday to Thursday), whereas the Friday and Saturday are two weekly holidays. Our investigation aims at predicting the current day's closing price based on previous days' values of different stocks while analyzing the EOD features. For weekly charts, current week's closing price is predicted based on previous weeks' values. The date feature is used for the purpose of grouping previous few days' or weeks' data together and the corresponding closing price is considered to be the ground truth label.

\begin{figure}[htbp]
    \centering
    \includegraphics[width=\textwidth, keepaspectratio]{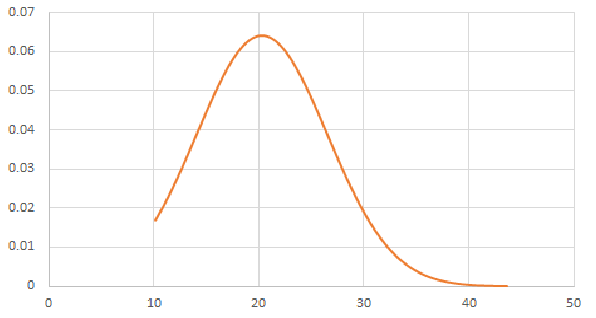}
    \caption{Distribution of the EOD closing price of a selected stock, namely AGRANINS}
    \label{fig:AGRANINSDataDistribution}
\end{figure}

A total of eight different companies are considered for the experimental studies, which are listed in Table~\ref{tab:companylist}. \textit{AAMRANET} and \textit{AGRANINS} are two of them. The data of these two companies are given in Table~\ref{tab:data_analysis_raw}. The distribution for closing price column in \textit{AGRANINS} daily chart data is shown in Figure~\ref{fig:AGRANINSDataDistribution}. Similar graph can be drawn for all the instances of all the companies. 

\begin{table}[htbp]
    \centering
    \caption{Analysis on some input data}
    \label{tab:data_analysis_raw}
    \resizebox{\textwidth}{!}{
    \begin{tabular}{cccccccccc}
    \hline
    \multirow{2}{*}{\textbf{Trading Code}}  &    \multirow{2}{*}{\textbf{Data Column}}  &   \multicolumn{4}{c}{\textbf{Daily Chart Data}} & \multicolumn{4}{c}{\textbf{Weekly Chart Data}} \\ \cline{3-10}
        &   &   \multicolumn{1}{c}{\textbf{Minimum}}   & \multicolumn{1}{c}{\textbf{Maximum}}   &   \multicolumn{1}{c}{\textbf{Mean}}    &   \multicolumn{1}{p{1.8cm}}{\textbf{Standard Deviation}}  &   \multicolumn{1}{c}{\textbf{Minimum}} &   \multicolumn{1}{c}{\textbf{Maximum}} &   \multicolumn{1}{c}{\textbf{Mean}}    &   \multicolumn{1}{p{1.8cm}}{\textbf{Standard Deviation}} \\ \hline
    \multirow{5}{*}{\textit{AAMRANET}}  &   Opening Price   &   32.2    &   136.8868 &   59.3027  & 22.2639   &   33.2    &   130 &   59.2921 &   21.9127 \\ 
        &   Highest Price   &   32.5    &   147.7358 &   60.4845 &   23.0455 &   33.2    &   147.7358 &   62.3512 &   23.7689 \\ 
        &   Lowest Price    &   28.3    &   134.9057 &   58.0309 &   21.6448 &   28.3    &   101.9811 &   56.2527 &   20.3377 \\ 
        &   Closing Price   &   28.9    &   136.3208 &   58.9793 &   22.2044 &   33.1    &   128.3019 &   58.8170 &   21.7957 \\ 
        & Volume    &   0   &   5689599 &   291662.655  &   440741.414  &   6830    &   13318623    &   1449416.318 &   1932115.901 \\ \hline
    \multirow{5}{*}{\textit{AGRANINS}}  &   Opening Price   &   9.9773 &   43.7    &    20.3345 &   6.3306 &   10.8844 &   41.2    &   20.2642 &   6.2596 \\ 
        &   Highest Price   &   10.5215 &   44.2    &   20.8126 &   6.5631 &   11.9728 &   44.2    &   21.5340 &   6.9405 \\ 
        &   Lowest Price    &   9.9773 &   41  &   19.8957  &   5.9832  &   9.9773 &   35.6    &   19.2381 &   5.5807    \\  
        &   Closing Price    &   10.15873   &   43.6    & 20.2915 & 6.20802 &   10.70295    & 41.2  & 2027062   &   6.192051    \\  
        &   Volume  &   0   &   5252869 &   181003.85   &   432506.4306 &   8996    &   13580904    &   903871.9299 &   1955935.685 \\ \hline
    \end{tabular}
    }
\end{table}

Stock related prices in raw usually vary quite a lot. For effective learning of a machine learning model, it is quite important for data to be stationary to a reasonable extent. To introduce the ``stationarity'' property, the raw feature values are converted into return values or differential values, which means that we subtract $i^{th}$ day\textquotesingle s (e.g., today\textquotesingle s) price from the $(i-1)^{th}$ day\textquotesingle s (e.g., yesterday\textquotesingle s)  price to compute the $i^{th}$ value of the stationary series. Also, before converting to stationary data, each value is transformed into the mean average of that day with a windows size of 10 for calculating the mean.

\begin{figure}[htbp]
\centerline{\includegraphics[width = 0.9\textwidth, keepaspectratio]{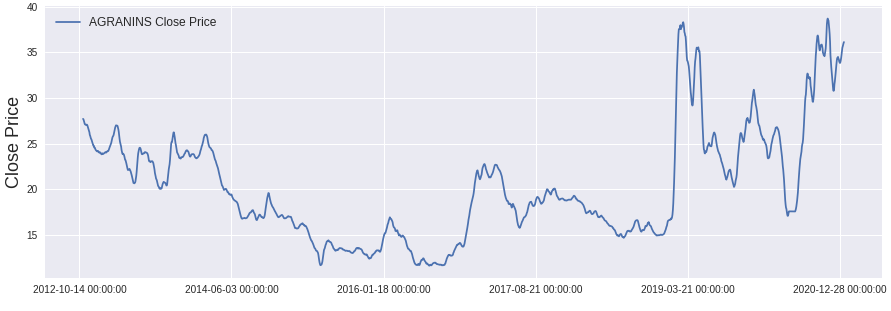}}
\caption{Closing price of a selected Stock, namely AGRANINS over a time window of around eight years}
\label{fig:AGRANINS_RawData}
\end{figure}

\begin{figure}[htbp]
\centerline{\includegraphics[width = 0.9\textwidth, keepaspectratio]{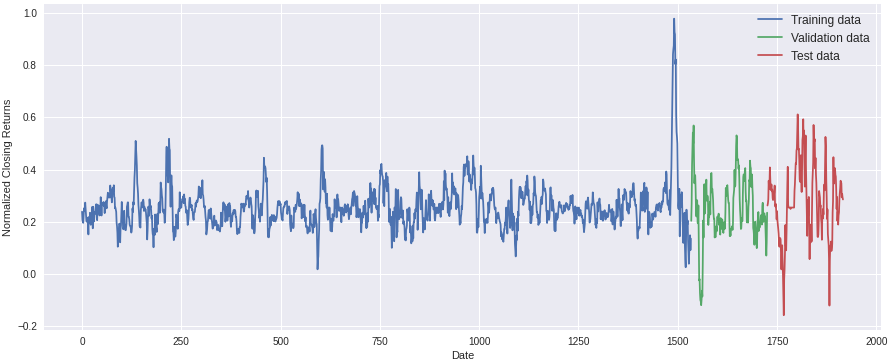}}
\caption{Dataset of a selected stock, namely  AGRANINS after introducing data stationarity and normalizations}
\label{fig:AGRANINS_SplittedNormalized}
\end{figure}

\begin{figure}[htbp]
    \centering
    \includegraphics[width = \textwidth, keepaspectratio]{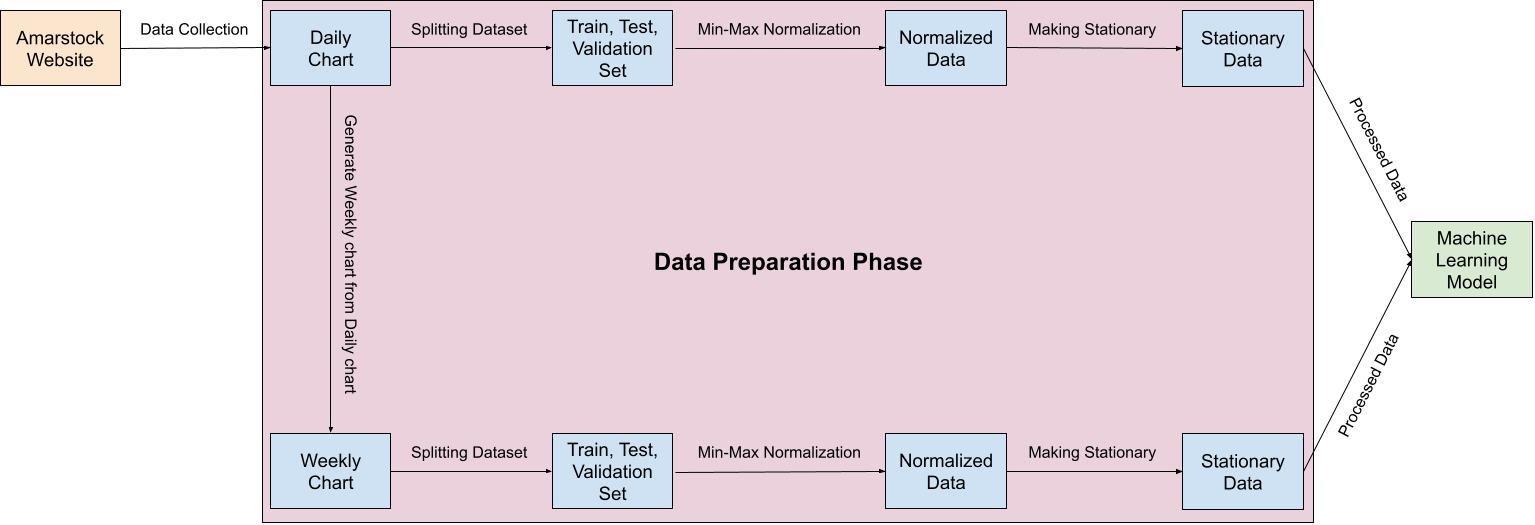}
    \caption{Steps in the data preparation stage before feeding into the machine learning model
    }
    \label{fig:data_prep_diagram}
\end{figure}

The entire data are normalized using Min-Max normalization technique. The raw closing data of one of the selected stocks, namely \textit{AGRANINS} is shown in Figure~\ref{fig:AGRANINS_RawData}. After making the data stationary, followed by normalization and splitting, the resulted data are shown in Figure~\ref{fig:AGRANINS_SplittedNormalized}. It is generally known that stock market raw data are non-stationary in nature, which makes them hard to be modelled and predicted. It can be easily observed from the Figure~\ref{fig:AGRANINS_RawData} and Figure~\ref{fig:AGRANINS_SplittedNormalized} that data of Figure~\ref{fig:AGRANINS_SplittedNormalized} have more stationarity property (i.e., the mean, variance and auto-correlation structure do not vary over time) than that of Figure~\ref{fig:AGRANINS_RawData}. Thus we expect to make better prediction after this preprocessing of stationarity and normalization than the original data. The entire data collection and preparation phase is demonstrated in a diagram presented in Figure~\ref{fig:data_prep_diagram}. 

After these preprocessing, each row of dataset contains stationary, normalized data of current day\textquotesingle s closing price return as the label and the opening price, highest price, lowest price, closing price and volume return data. For the prediction of closing price of a day (week), the previous eight days\textquotesingle (weeks\textquotesingle) data are taken into account during the training of the transformer model. 

\begin{table}[htbp]
    \centering
    \caption{Eight companies selected for modeling and prediction}
    \begin{tabular}{lr}
    \hline
    \textbf{Company Name} & \textbf{Trading Code} \\
    \hline \hline
    First Janata Bank Mutual Fund & 1JANATAMF \\
    Aamra Networks Limited & AAMRANET \\
    AB Bank Limited & ABBANK \\
    ACI Limited & ACI \\
    ACI Formulations Limited & ACIFORMULA \\
    Agrani Insurance Company Limited & AGRANINS \\
    Alltex Industries Limited & ALLTEX \\
    Delta Life Insurance Company Limited & DELTALIFE \\ \hline
    \end{tabular}
    \label{tab:companylist}
\end{table}

\subsection{Transformer-Based Learning Model with \emph{time2vec} Encoding}
The encoding of input data is a crucial part for the neural networks. Since our data are time series, i.e., prediction of a feature vector depends on the previously seen feature vectors, we use the recently introduced \emph{time2vec} scheme to encode our data. The input of the model consists of 5 features from 8 days and thus its size becomes $(\textit{batch\_size}, 8, 5)$. We take 32 as the batch size and so the input shape consists of the shape $(32, 8, 5)$. 

\begin{figure*}[htbp]
\centerline{\includegraphics[width = 0.9\textwidth, height = 0.9\textheight, keepaspectratio]{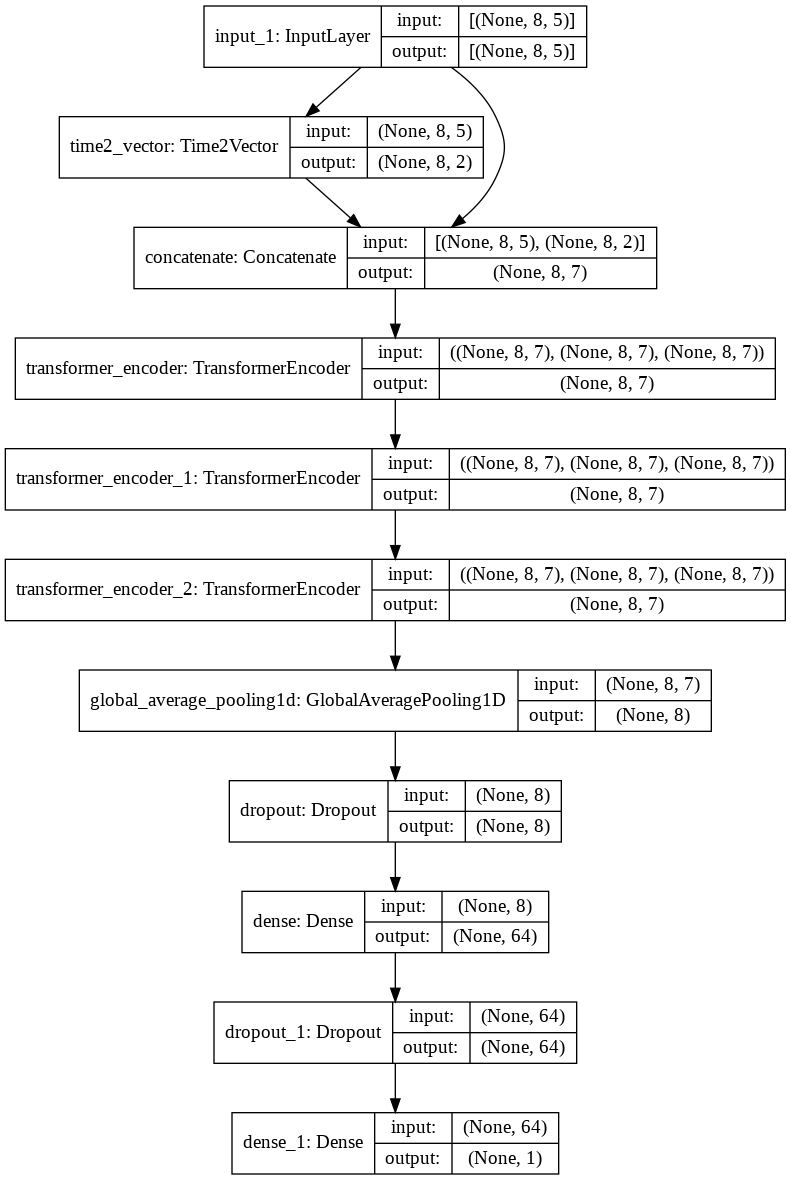}}
\caption{Architecture of the proposed transformer-based framework}
\label{fig:architecture}
\end{figure*}

The self-attention mechanism \cite{vaswani2017attention-transformer} is a clever idea that emphasizes on specific parts of the input sequence for making prediction. In our model, this idea is established using single and multi-headed attention. Some dense layers and a global average pooling layer are also added to the model to effectively decode the encoded input and to make the model robust respectively. Since the study attempts to predict specific value of the stock price, the transformer model output consists of only a single neuron that provides a continuous value. The designed model employs only encoder, but no decoder, and thus it is quite similar to popular encoder models like BERT architecture \cite{devlin2018bert}. However, unlike BERT, the proposed model incorporates time embedding which should give it a performance boost when it comes to time series prediction problems such as, in our case, stock price prediction. The whole model architecture is presented in Figure~\ref{fig:architecture}.

With the daily chart data, a total of eight different transformer based models have been trained (i.e., one for each of the eight stocks mentioned in Table~\ref{tab:companylist}). The training and validation phases are conducted throughout 50 epochs, and the error chart shown in Figure~\ref{fig:AAMRANET_Error} is found to be starting to flatten at that point. After these phases, the trained models are evaluated using the test data.\footnote{The data are split 8:1:1 for training : validation : testing} Since the test data have never been used during the training phase, the results demonstrate how well the model can perform on unseen data, which is an indication of the generalization capability of the trained models.

Once completed, the same procedure is followed for the weekly chart data as well. The same eight companies' weekly data are used to train and validate eight more models, which are then tested following the same procedure. The complete experiment thus consist of 16 transformer models. The combination is shown in Table~\ref{tab:model_combo}.

\begin{table}[htbp]
    \centering
    \caption{Prediction models developed in the study}
    \begin{tabular}{ccc}
    \hline
    \textbf{Chart Type} & \textbf{Company Name} & \textbf{Model No.} \\ \hline \hline
        \multirow{8}{4em}{Daily} & 1JANATAMF & Model 01 \\ 
         & AAMRANET & Model 02 \\ 
         & ABBANK & Model 03 \\ 
         & ACI & Model 04 \\ 
         & ACIFORMULA & Model 05 \\ 
         & AGRANINS & Model 06 \\ 
         & ALLTEX & Model 07 \\ 
         & DELTALIFE & Model 08 \\ \hline
         \multirow{8}{4em}{Weekly} & 1JANATAMF & Model 09 \\ 
         & AAMRANET & Model 10 \\ 
         & ABBANK & Model 11 \\ 
         & ACI & Model 12 \\ 
         & ACIFORMULA & Model 13 \\ 
         & AGRANINS & Model 14 \\ 
         & ALLTEX & Model 15 \\ 
         & DELTALIFE & Model 16 \\ \hline
    \end{tabular}
    \label{tab:model_combo}
\end{table}

\section{Result Analysis}
\label{sec: result}
The proposed model predicts closing price returns of each of the selected stocks. From closing returns, the actual closing price can be generated. In the daily chart based models, the closing price of $t^{th}$ day is predicted using the values of previous eight days, i.e., using the values of $(t-1)^{th}, (t-2)^{th}, ..., (t-8)^{th}$ days. Similarly, for the weekly chart based models, the closing price of $t^{th}$ week is predicted using the values of previous wight weeks.

\begin{figure}[htbp]
    \centerline{\includegraphics[width = .9\textwidth, keepaspectratio]{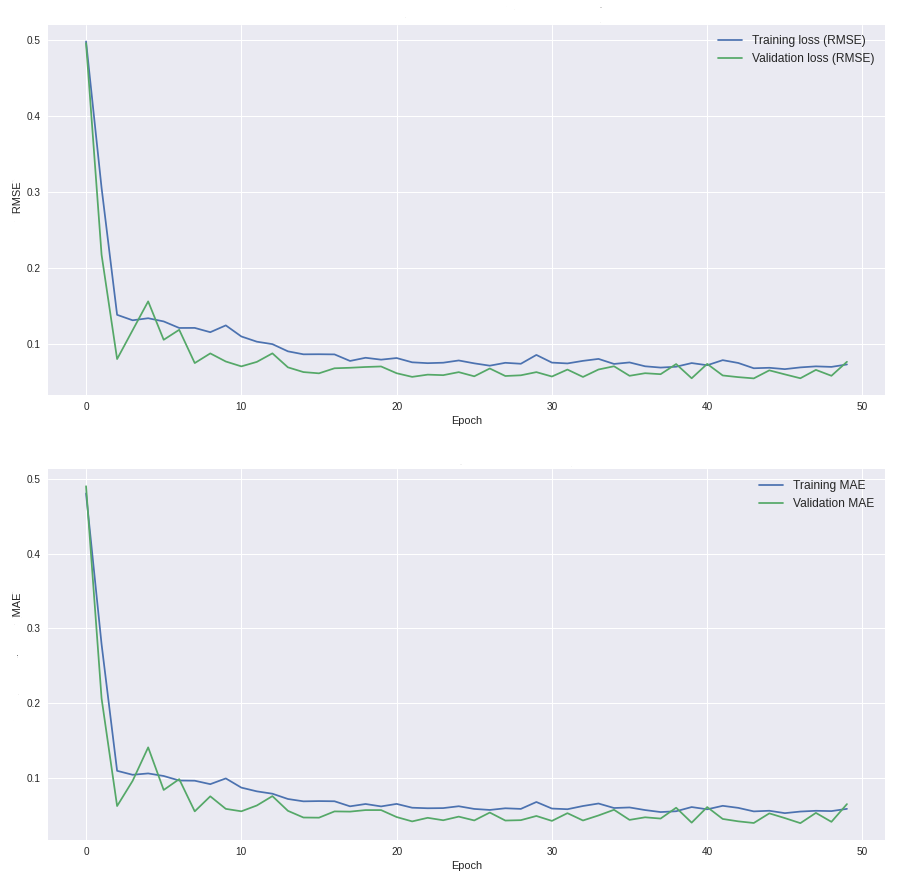}}
    \caption{Training errors of proposed transformer-based model for the stock AAMRANET}
    \label{fig:AAMRANET_Error}
\end{figure}

\begin{figure}[htbp]
    \centering
    \begin{subfigure}{\textwidth}
        \centering
        \includegraphics[width = 0.9\textwidth, keepaspectratio]{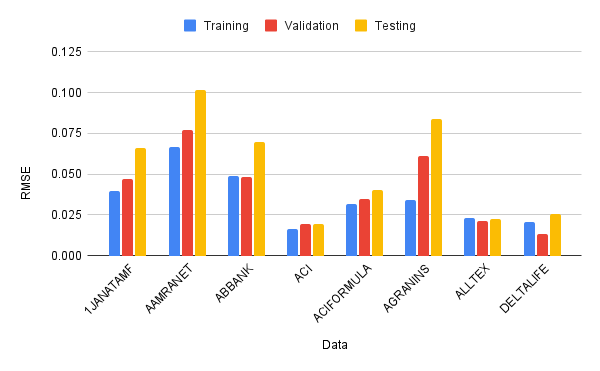}
        \caption{RMSE value for each stock}
        \label{fig:daily_error_comp_rmse}
    \end{subfigure}
    \begin{subfigure}{\textwidth}
        \centering
        \includegraphics[width = 0.9\textwidth, keepaspectratio]{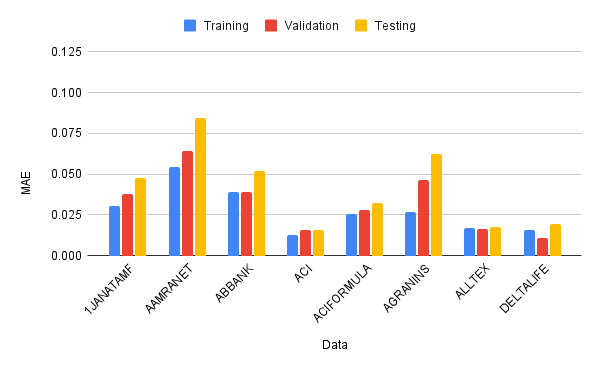}
        \caption{MAE value for each stock}
        \label{fig:daily_error_comp_mae}
    \end{subfigure}
    \caption{Comparison of error values of the selected eight stocks by the daily prediction model}
    \label{fig:daily_error_comp}
\end{figure}

\begin{figure}[htbp]
    \centering
    \begin{subfigure}{\textwidth}
        \centering
        \includegraphics[width = 0.9\textwidth, keepaspectratio]{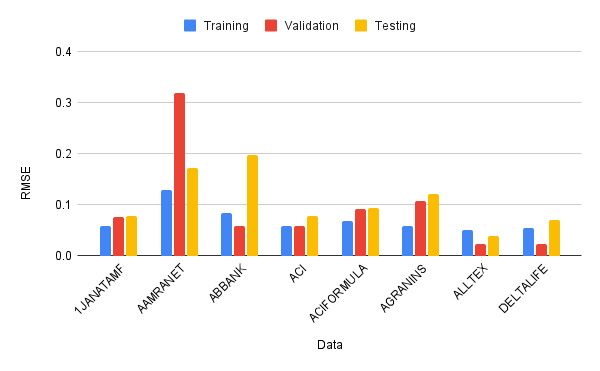}
        \caption{RMSE value for each stock}
        \label{fig:weekly_error_comp_rmse}
    \end{subfigure}
    \begin{subfigure}{\textwidth}
        \centering
        \includegraphics[width = 0.9\textwidth, keepaspectratio]{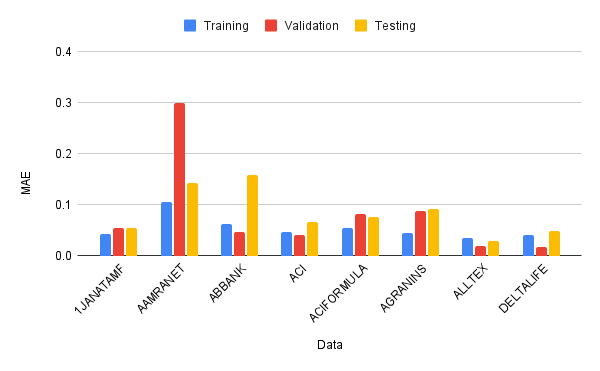}
        \caption{MAE value for each stock}
        \label{fig:weekly_error_comp_mae}
    \end{subfigure}
    \caption{Comparison of error values of the selected eight stocks by the weekly prediction model}
    \label{fig:weekly_error_comp}
\end{figure}

\begin{figure}[htbp]
    \centerline{\includegraphics[width = 0.9\textwidth, keepaspectratio]{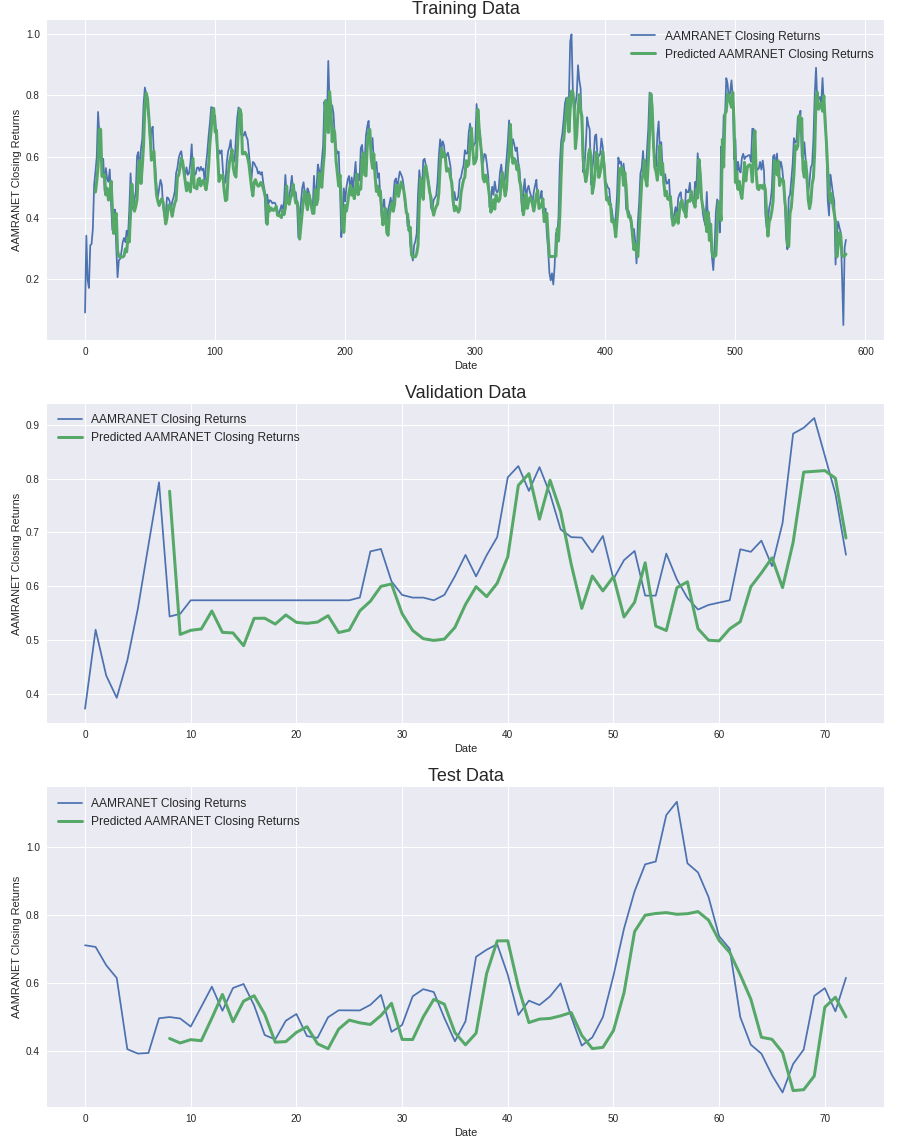}}
    \caption{Prediction performance of trained model on AAMRANET stock using the processed data}
    \label{fig:AAMRANET_ProcessedPrediction}
\end{figure}

\begin{figure}[htbp]
    \centerline{\includegraphics[width = 0.9\textwidth, keepaspectratio]{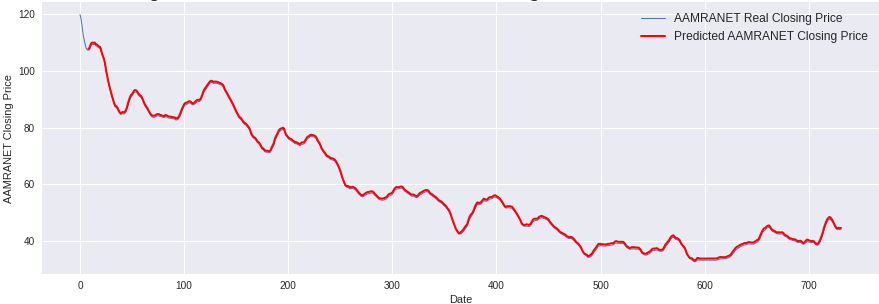}}
    \caption{Prediction performance of trained model on AAMRANET stock using real raw data}
    \label{fig:AAMRANET_FullPrediction}
\end{figure}

The results shown in Table~\ref{tab:errorvals} indicate that the proposed transformer based prediction models can make satisfactory price prediction with low error. The performance on a particular stock --- AAMRANET (daily chart) is presented as a representative here. Figure~\ref{fig:AAMRANET_Error} shows the error value (RMSE and MAE) per epoch. A more succinct visualization of errors of all the conducted experiments can be seen in Figures~\ref{fig:daily_error_comp} and \ref{fig:weekly_error_comp}. Figure~\ref{fig:AAMRANET_ProcessedPrediction} presents the prediction of closing price returns. Finally, Figure~\ref{fig:AAMRANET_FullPrediction} depicts the prediction of real closing price of the stock. 

The model was trained with an aim to reduce the root mean square (RMSE) error value. Table~\ref{tab:errorvals} presents the error values of the trained models on each of selected eight stocks of DSE.

\begin{table}[htbp]
\caption{Error values of the transformer-based proposed model on eight stocks from DSE using both daily and weekly data}
\begin{center}
\resizebox{\textwidth}{!}{\begin{tabular}{ccccccccccc}
\hline
\textbf{Chart Type} & \textbf{Data} & \textbf{Error Name} & \textbf{1JANATAMF} & \textbf{AAMRANET} & \textbf{ABBANK} & \textbf{ACI} & \textbf{ACIFORMULA} & \textbf{AGRANINS} & \textbf{ALLTEX} & \textbf{DELTALIFE} \\
\hline \hline
\multirow{6}{4em}{Daily} & \multirow{2}{4.5em}{Training} & RMSE & 3.94E-02 & 6.69E-02 & 4.90E-02 & 1.62E-02 & 3.18E-02 & 3.39E-02 & 2.31E-02 & 2.05E-02 \\ 
 & & MAE & 3.07E-02 & 5.45E-02 & 3.94E-02 & 1.25E-02 & 2.54E-02 & 2.66E-02 & 1.71E-02 & 1.61E-02 \\ \cline{2-11}
 & \multirow{2}{4.5em}{Validation} & RMSE & 4.68E-02 & 7.71E-02 & 4.84E-02 & 1.95E-02 & 3.48E-02 & 6.09E-02 & 2.11E-02 & 1.33E-02 \\ 
 & & MAE & 3.77E-02 & 6.45E-02 & 3.90E-02 & 1.55E-02 & 2.79E-02 & 4.67E-02 & 1.62E-02 & 1.08E-02 \\ \cline{2-11}
 & \multirow{2}{4.5em}{Testing} & RMSE & 6.59E-02 & 1.02E-01 & 6.96E-02 & 1.93E-02 & 4.06E-02 & 8.39E-02 & 2.23E-02 & 2.54E-02 \\ 
 & & MAE & 4.76E-02 & 8.43E-02 & 5.20E-02 & 1.55E-02 & 3.23E-02 & 6.25E-02 & 1.74E-02 & 1.94E-02 \\ \hline
\multirow{6}{4em}{Weekly} & \multirow{2}{4.5em}{Training} & RMSE & 5.78E-02 & 1.30E-01 & 8.35E-02 & 5.83E-02 & 6.82E-02 & 5.78E-02 & 5.05E-02 & 5.46E-02 \\ 
 & & MAE & 4.33E-02 & 1.06E-01 & 6.32E-02 & 4.73E-02 & 5.39E-02 & 4.54E-02 & 3.52E-02 & 4.05E-02 \\ \cline{2-11}
 & \multirow{2}{4.5em}{Validation} & RMSE & 7.54E-02 & 3.19E-01 & 5.90E-02 & 5.90E-02 & 9.13E-02 & 1.07E-01 & 2.38E-02 & 2.27E-02 \\ 
 & & MAE & 5.49E-02 & 3.00E-01 & 4.65E-02 & 3.98E-02 & 8.24E-02 & 8.79E-02 & 2.00E-02 & 1.79E-02 \\ \cline{2-11}
 & \multirow{2}{4.5em}{Testing} & RMSE & 7.70E-02 & 1.72E-01 & 1.98E-01 & 7.79E-02 & 9.31E-02 & 1.21E-01 & 3.87E-02 & 6.95E-02 \\ 
 & & MAE & 5.53E-02 & 1.42E-01 & 1.59E-01 & 6.55E-02 & 7.64E-02 & 9.09E-02 & 2.98E-02 & 4.80E-02 \\ \hline
\end{tabular}}
\label{tab:errorvals}
\end{center}
\end{table}

\begin{figure}
    \centering
    \begin{subfigure}{\textwidth}
        \centering
        \includegraphics[height=0.1\textheight, width=0.9\textwidth]{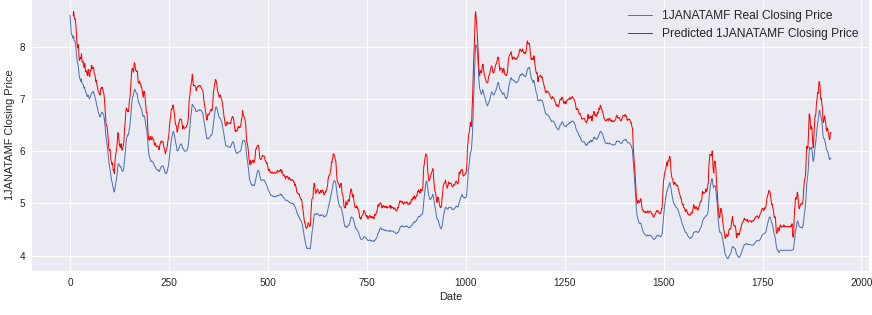}
        \caption{1JANATAMF}
        \label{fig:1JANATAMF_Full}
    \end{subfigure}
    \begin{subfigure}{\textwidth}
        \centering
        \includegraphics[height=0.1\textheight, width=0.9\textwidth]{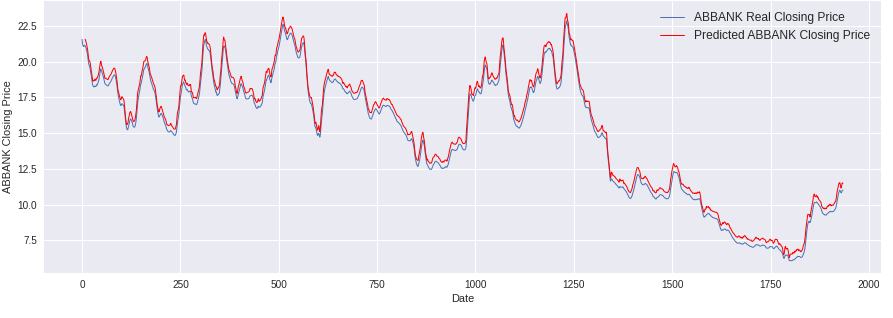}
        \caption{ABBANK}
        \label{fig:ABBANK_Full}
    \end{subfigure}
    \begin{subfigure}{\textwidth}
        \centering
        \includegraphics[height=0.1\textheight, width=0.9\textwidth]{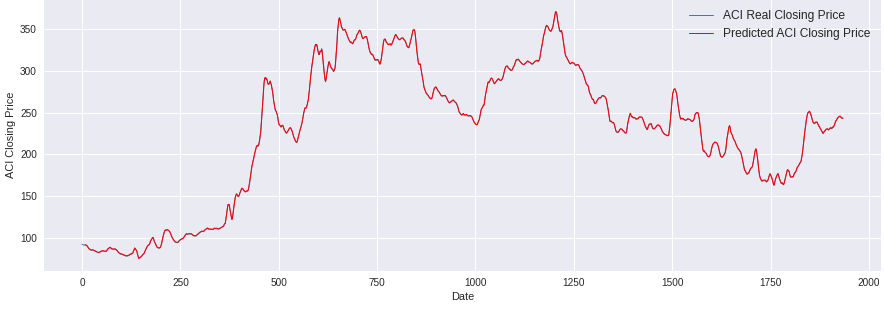}
        \caption{ACI}
        \label{fig:ACI_Full}
    \end{subfigure}
    \begin{subfigure}{\textwidth}
        \centering
        \includegraphics[height=0.1\textheight, width=0.9\textwidth]{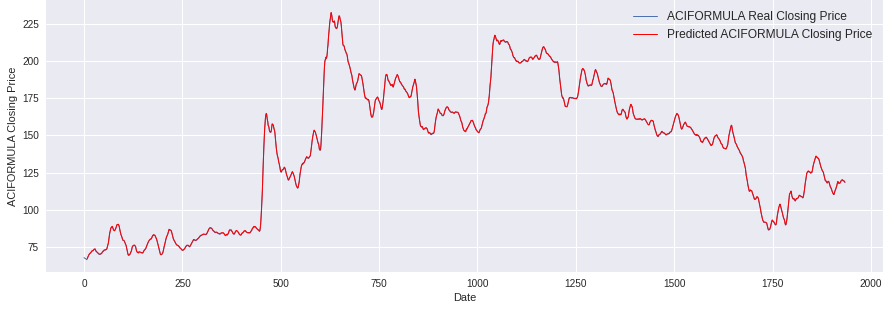}
        \caption{ACIFORMULA}
        \label{fig:ACIFORMULA_Full}
    \end{subfigure}
    \begin{subfigure}{\textwidth}
        \centering
        \includegraphics[height=0.1\textheight, width=0.9\textwidth]{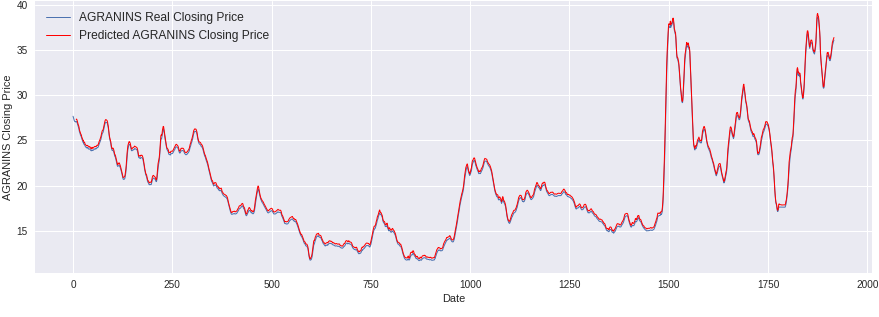}
        \caption{AGRANINS}
        \label{fig:AGRANINS_Full}
    \end{subfigure}
    \begin{subfigure}{\textwidth}
        \centering
        \includegraphics[height=0.1\textheight, width=0.9\textwidth]{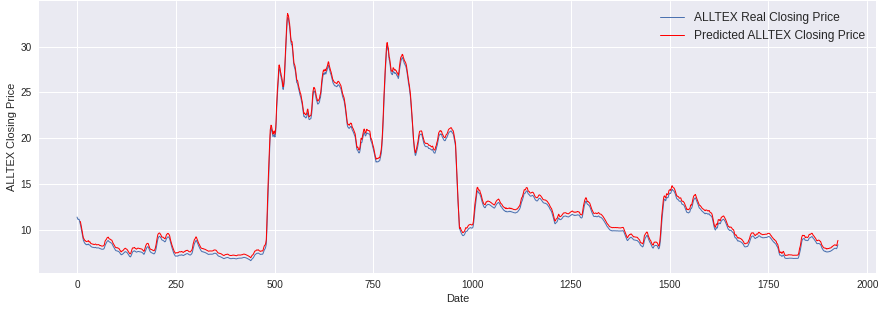}
        \caption{ALLTEX}
        \label{fig:ALLTEX_Full}
    \end{subfigure}
    \begin{subfigure}{\textwidth}
        \centering
        \includegraphics[height=0.1\textheight, width=0.9\textwidth]{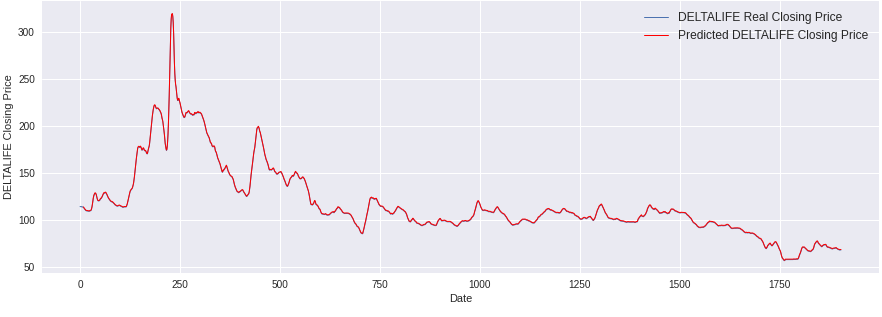}
        \caption{DELTALIFE}
        \label{fig:DELTALIFE_Full}
    \end{subfigure}
    \caption{Prediction performance of trained model on seven different stocks in real raw data}
    \label{fig:seven_preds}
\end{figure}

\begin{figure}
    \centering
    \includegraphics[width = 0.9\textwidth, height = 0.45\textheight]{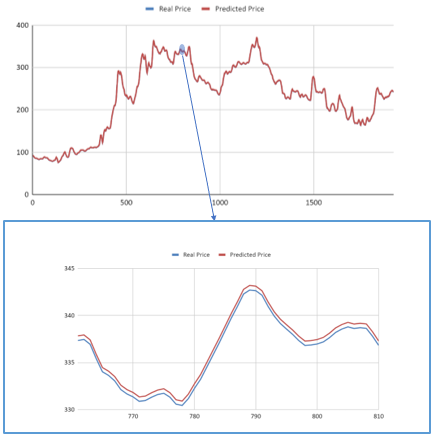}
    \caption{Zoomed image showing the difference of predicted value and real value for \textit{ACI}}
    \label{fig:ACI_Zoomed}
\end{figure}

\begin{figure}
    \centering
    \includegraphics[width = 0.9\textwidth, height = 0.45\textheight]{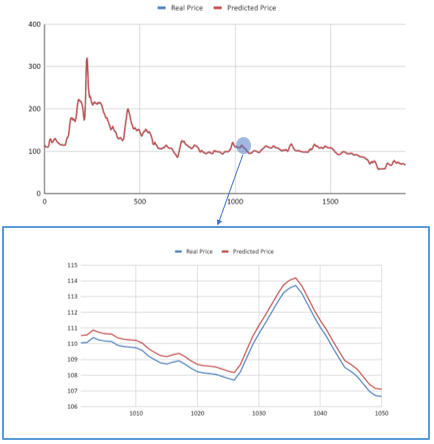}
    \caption{Zoomed image showing the difference of predicted value and real value for \textit{DELTALIFE}}
    \label{fig:DELTALIFE_Zoomed}
\end{figure}

\begin{figure}
    \centering
    \includegraphics[width=\textwidth, keepaspectratio]{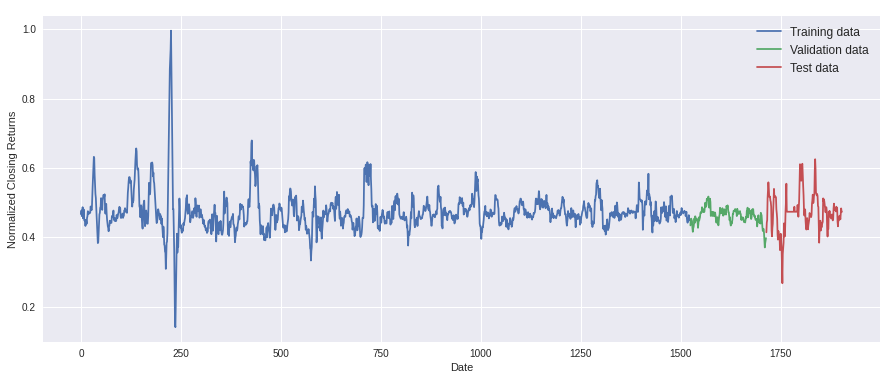}
    \caption{DELTALIFE data after making raw data stationary and applying normalization}
    \label{fig:deltalife_processed}
\end{figure}

Figure~\ref{fig:seven_preds} presents the visual comparison between predicted price and actual closing price of the remaining seven  stocks. From the comparative chart plots, it is apparent that most of the predictions are very close to the actual price, and the predictions often overlap with the actual price. The predictions seem little more deviated for two stocks --- \textit{1JANATAMF} and \textit{ABBANK}. All the other companies' predicted and real data are not much different, as shown in Figures~\ref{fig:AAMRANET_FullPrediction} and ~\ref{fig:seven_preds}. Actually, in some of the charts only one plot is visible (e.g., \textit{ACI}, \textit{ACIFORMULA}, \textit{DELTALIFE}), because the predicted value almost overlays on the actual value and their difference is only visible in the zoomed figure, as shown in Figures ~\ref{fig:ACI_Zoomed} and ~\ref{fig:DELTALIFE_Zoomed}. Another observation (Table~\ref{tab:errorvals}) is that the training dataset error is higher than the validation and testing datasets. After some analysis, the reason behind this is found to be the characteristics of the dataset itself --- the presence or absence of spikes, peaks and valleys in training, validation and test sets. For example, the training error for \textit{DELTALIFE} is higher than the validation error, and the reason becomes visible in the Figure~\ref{fig:deltalife_processed}, which clearly shows that there exist several high peaks and valleys in the training set, which are not present in the validation set. Similar characteristics lead to relatively higher training error and lower validation and/or test errors.

\section{Conclusion}
\label{sec: conclusion}

The primary contribution of this paper is to introduce the application of a transformer based machine learning model for predicting the future prices of stock listed in Dhaka Stock Exchange (DSE). A similar approach may be designed for generic time series analysis and signal processing problems. The experimental results show that such models can achieve satisfactory low error rates, even within a short execution time. The proposed models can predict future price not only after a trading day, but also after a week or a month. In other words, the benefits of transformer based models can be harnessed for taking both short and long time investment decision. Such a model can come to a great help for an investor to understand, analyze and predict the market movement. 

Although the proposed model addresses the price prediction problem as a regression problem, it can be easily modified to deal with classification problem. For example, a model can be designed to predict whether the price will rise or fall in the upcoming days (thereby dealing with a binary classification problem). In short, transformer based machine learning models are versatile and flexible enough to deal with diverse types of problems like these, and thus present bright possibility to assist the general investors most of whom usually invest in the stock market to earn their very livelihood. Such positive outcome might be more visible for the investors of the stock market of an underdeveloped or developing country like Bangladesh, which was the focus of this study.

\bibliographystyle{unsrt}  
\bibliography{arXiv}  






\end{document}